\documentclass[symmetry,article,accept,oneauthor,pdftex]{Definitions/mdpi}

\usepackage{color}

\firstpage{1} 
\pubvolume{14}
\issuenum{2}
\articlenumber{300}
\pubyear{2022}
\copyrightyear{2022}
\externaleditor{{Academic Editor: Maxim Y. Khlopov} %MDPI: please add this part.
} % For journal Automation, please change Academic Editor to "Communicated by"
\datereceived{23 December 2021} 
\dateaccepted{28 January 2022} 
\datepublished{1 Feb 2022} 
\hreflink{https://www.mdpi.com/\linebreak 2073-8994/14/2/300} % If needed use \linebreak
\def\ba{\begin{array}}
\def\ea{\end{array}} 
\def\bea{\begin{eqnarray}}
\def\eea{\end{eqnarray}}
\def\beq{\begin{equation}}
\def\eeq{\end{equation}}
\def\ben{\begin{enumerate}}
\def\een{\end{enumerate}}

\def\brr{\begin{array}}
\def\err{\end{array}}

\def\calMa{{V_{4}}}

\def\rL{r_\Lambda}

\def\dA{{d\Omega}}

\Title{The Cosmological Constant as Event Horizon}
\TitleCitation{The Cosmological Constant as
Event Horizon}
\Author{{Enrique Gaztanaga} %Please carefully check the accuracy of names and affiliations.
 $^{1,2}$\orcidA{}}

\AuthorNames{Enrique Gaztanaga}

\AuthorCitation{Gaztanaga, E.}
\address{%
$^{1}$ \quad Institute of Space Sciences (ICE, CSIC), 
08193 Barcelona, Spain; {gaztanaga@darkcosmos.com} 
 \\
$^{2}$ \quad Institut d Estudis Espacials de Catalunya (IEEC), 08034 Barcelona,  Spain}

\abstract{General Relativity allows for a cosmological constant ($\Lambda$) which has inspired models of cosmic Inflation and Dark Energy. We show instead that $r_\Lambda = \sqrt{3/\Lambda}$ corresponds to an event horizon: a causal boundary term in the action. Our Universe is expanding inside its Schwarzschild radius $r_S=\rL=2GM$,  which could have originated from a uniform free falling cloud of mass $M$ that collapsed as a Black Hole (BH) 25 Gyrs ago. Such a BH Universe allows for large-scale structure formation without the need of Inflation or Dark Energy.}


\begin{document}
%% \linenumbers

%% main text
\section{Introduction}
\label{S:1}

The cosmological constant $\Lambda$ has a very interesting history. 
For~a review,
 see~\cite{Weinberg1989,Carroll1992,Peebles2003}.
In 1917, right after the first publications of General Relativity (GR) \cite{Einstein1916}, Albert Einstein applied GR to the Universe as a whole~\cite{Einstein1917}. This was before the discovery of cosmic expansion~\cite{Elizalde2021}. Einstein realized that to account for the whole Universe, we need to include  a cosmological constant, $\Lambda$, as~ a new universal constant of nature.  
At~that time,
 most scientists believed that the Universe has to be in static equilibrium on large scales. 
 This is what is usually assumed for large physical systems that are allowed to evolve in isolation for a long time.  %PLEASE CHECK.
 However,~gravity is always attractive and it is therefore hard to imagine how you can reach such an equilibrium on the largest scales where other forces are negligible. 
 This is why Einstein had to invoke $\Lambda$.

We can see this by considering  the GR version of the Poisson equation for a perfect fluid with density $\rho$ and pressure $p$
(see Equation~(12) in~\cite{Gaztanaga2021}):
\beq
\nabla_\mu 
{\text{g}^\mu} =
\frac{d\Theta}{d\tau} + \frac{1}{3} \Theta^2 = R_{\mu\nu} u^\mu u^\nu 
= \Lambda - 4\pi G(\rho+3p) 
 \label{eq:rPoisson} 
\eeq
where ${\text{g}^\mu}$ is the geodesic acceleration~\cite{Padmanabhan}. 
This is also the Raychaudhuri equation for a shear-free, non-rotating fluid where $\Theta = \nabla_\nu u^\nu$ and $u^\nu$ is the  4-velocity.
The above equation is purely geometric: it describes the evolution in proper time $\tau$ of the dilatation coefficient $\Theta$ of a bundle of nearby geodesics.
Note that without $\Lambda$, the acceleration is always negative and there is no way to reach an equilibrium, unless~$p<-1/3 \rho$ which is what we call Dark Energy~today.

The $\Lambda$ term in the above equation is equivalent to a vacuum or a ground state of constant density (and negative pressure). It also corresponds to a centrifugal (or Hooke's) force~\cite{CalderLahav2008,Gaztanaga2020}
in the Newtonian potential (or metric element): $2\Phi= -\Lambda r^2/3$ that opposes Newtonian gravity,  $2\Phi= -2GM/r$. Equilibrium happens when both forces are equal, similar to the circular Kepler orbits in Newtonian~dynamics. 

The $\Lambda$ term was mostly abandoned after the Hubble--Lemaitre expansion law was accepted, which took some time~\cite{Elizalde2021}. 
In~1931, an~unpublished manuscript of Einstein~\cite{SteadyState}  considered the possibility of a Universe that expands with $\Lambda$ but remains static due to a continuous formation of matter from empty space, as~in the  Steady State \mbox{Cosmology~(\cite{SSC1,SSC2}).}
This has the advantage of a Perfect Cosmological Principle, where the universe is homogeneous, isotropic and static. This is in better agreement with the relativity principle than the regular Cosmological Principle, which only applies to space. 
The~key to this idea is the static deSitter (dS) metric:
  \beq
2\Phi= 2\Psi = - r^2/r_*^2
\label{eq:deSitter}
\eeq
where $r_*$ is a constant that we will identify later 
as $r_*=r_\Lambda \equiv 1/H_\Lambda \equiv \sqrt{\Lambda/3}$ 
in our Universe.
$\Phi$ and $\Psi$ are the gravitational potentials of  a generic metric with spherical symmetry in physical or Schwarzschild (SW) coordinates $(t,r,\theta, \varphi)$ \cite{Padmanabhan,Dodelson}:
\beq
 ds^2 = g_{\mu\nu} dx^\mu dx^\nu =
 -(1+2\Psi) dt^2 +  \frac{dr^2}{1+2\Phi} + r^2 \dA^2
\label{eq:newFRW}
 \eeq 
 where we have introduced the solid angle: 
$\dA^2 = d\theta^2 + \sin{\theta}^2 d\varphi^2$.

The idea of the  Steady State Cosmology solution was rejected because of the lack of observational evidence for continuous matter creation (to compensate the dilution by the expansion). 
However,~this was a misunderstanding. There is no need for ad~hoc matter creation to reconcile the observed cosmic expansion with $\Lambda$. The~frame duality~\cite{hal-03344159,universe}:
 \beq
 -d\tau^2 + a^2(\tau) d\chi^2 + r^2 \dA^2 = -(1+2\Psi) dt^2 +  \frac{dr^2}{1- r^2 H^2} + r^2 \dA^2 
% - d\tau^2 + a^2(\tau) d\chi^2  
\label{eq:newFR3}
 \eeq 
shows how the Friedmann–Lemaitre–Robertson–Walker (FLRW) in comoving coordinates
$(\tau,\chi,\theta,\varphi)$  (for an observer comoving with the  Hubble--Lemaitre expansion law) is the same time a quasi-static Universe in the SW frame. 
This is similar to the dS metric but with $H=H(\tau)=H(t,r)$. 
Under~this duality, the~Steady State Cosmology is not so different from the Big Bang FLRW expansion with $\Lambda$. Indeed, our FLRW metric (to the left of Equation~(\ref{eq:newFR3}) above) asymptotically approaches the dS metric in Equation~(\ref{eq:deSitter}). 
As~$H$ becomes dominated by $\Lambda$,
  the expansion becomes static  (see also~\cite{Mitra2012}) in the SW~frame.

Regardless of these considerations, it turns out that we need a $\Lambda$ term again today to explain the latest measurements of cosmic accelerated expansion~\cite{P18cosmo,DES2021}. An~effective $\Lambda$ term, similar to Dark Energy, might also be neeeded to understand cosmic \mbox{Inflation (\cite{Starobinski1979,Guth1981,Linde1982,Albrecht1982})}. The~Big Bang, Dark Energy, $\Lambda$, Inflation and BHs are puzzles we do not yet understand at any fundamental level. 
Are they somehow connected? 
What is the meaning of the measured $\Lambda$?

%%%%%%%%%%%%%%%%%%%%%%%%%%%%%%%%%%%%%%%%%%%%%
%%%%%%%%%%%%%%%%%%%%%%%%%%%%%%%%%%%%%%%%%%%%%
\section{The Action of GR and the \boldmath{$\Lambda$} term}

Consider the Einstein--Hilbert action (\cite{Hilbert1915,Padmanabhan}):
\beq
S = \int_\calMa d\calMa \left[ \frac{ R-2\Lambda}{16\pi G} +  {\cal L} \right],
\label{eq:action}
\eeq
where $d\calMa=\sqrt{-g} d^4 x$ is the invariant volume element, $\calMa$ is the volume of the 4D spacetime manifold, $R= R^\mu_\mu = g^{\mu\nu} R_{\mu\nu}$ is the Ricci scalar curvature and ${\cal L}$ the Lagrangian of the energy--matter content. We can obtain Einstein's field equations (EFE) for the metric field $g_{\mu\nu}$  from this action by requiring $S$ to be stationary $\delta S=0$ under arbitrary variations of the metric $\delta g^{\mu\nu}$. The~solution  is (\cite{Einstein1916,Padmanabhan}):

\beq
 G_{\mu\nu}+\Lambda g_{\mu\nu}=
 8\pi G~T_{\mu\nu} \equiv - \frac{16\pi G}{\sqrt{-g}} \frac{\delta (\sqrt{-g} {\cal L}) }{\delta g^{\mu\nu}},
\label{eq:rmunu}
\eeq
where $G_{\mu\nu} \equiv  R_{\mu\nu} - \frac{1}{2} g_{\mu\nu} R $.
For a perfect fluid in spherical coordinates:
\beq
T_{\mu\nu} =  (\rho+p) u_\mu u_\nu + p g_{\mu\nu}   
\label{eq:Tmunu}
\eeq 
where $\rho$,
and $p$ are the energy--matter density and pressure.
This fluid can be made of several components, each with a different equation of state $p=\omega \rho$.
For such a perfect fluid,
 the solution that minimizes the action (called action on-shell $S^{on-sh}$) results in a boundary term~\cite{Gaztanaga2021}:
\vspace{12pt}\beq
S^{on-sh} =
\int_\calMa d\calMa \frac{ \nabla_\mu {\textrm{g}^\mu} }{8\pi G}  =
\oint_{\partial \calMa}  \frac{ dV_\mu {\textrm{g}^\mu}}{8\pi G} 
=   \, <\frac{\Lambda}{4\pi G}-(\rho+3p)>_\calMa
\label{eq:S-on-sh}
\eeq
where, in the last step, we have used the Raychaudhuri equation (Equation~\eqref{eq:rPoisson}).
This is the situation when the evolution happens inside a BH. 
We want this boundary term to vanish (see below), so that its contribution to the action on-shell is zero: $S^{on-sh} = 0$ \cite{Gaztanaga2021}.  %PLEASE CHECK.
%For $\partial \calMa =\infty$ this means $\Lambda=0$. 
The observational fact that $\Lambda \ne 0$ implies that we live inside a boundary and to cancel the boundary term  we need: 
\beq
\Lambda =4\pi G <(\rho+3p)>_\calMa
\eeq
which basically tells us that $\Lambda$ is just given by the matter content inside such boundary. 
This is similar to what happens with the event horizon of a BH.  %PLEASE CHECK.
This gives an answer to the so-called coincidence problem~\cite{Peebles2003}: the coincidence between the value of $\rho_\Lambda \equiv \Lambda/(8\pi G)$ and $\rho$ matter today~\cite{Gaztanaga2020,Gaztanaga2021}.

Another way to see this is to note that 
Equation~(\ref{eq:rmunu}) requires that boundary terms vanish (e.g., see~\cite{Landau1971,Carroll2004,Padmanabhan}). 
Otherwise, we need to add a Gibbons--Hawking--York (GHY) boundary term~\cite{York,Gibbons,Hawking1996} to the action:

\beq
S = \int_\calMa d\calMa \left[ \frac{ R-2\Lambda}{16\pi G} +  {\cal L} \right]+ \frac{1}{8\pi G} \oint_{\partial \calMa}   d^3y \sqrt{-h} \,K. 
\label{eq:actionK}
\eeq
where $K$ is the trace of the extrinsic curvature at the boundary $\partial \calMa$ and $h$ is the induced metric. 
We will show explicitly in  Section~\ref{sec:GHY} that the GHY boundary  results in a $\Lambda$ term 
when the evolution happens following an FLRW metric inside an expanding BH event horizon. To~cancel the GHY term,
 we need $\rL=r_S$. 
 That $\Lambda$ is a GHY term was first proposed in~\cite{Gaztanaga2021} and has also been later interpreted as a boundary entropy term by~\cite{GB2021a,GB2021b,GB2021c}.

%%%%%%%%%%%%%%%%%%%%%%%%%%%%
\subsection{The Black Hole Universe (BHU) Solution}
\label{sec:BH.u}

Observations show that the expansion rate today is dominated by $\Lambda $. This indicates that the FLRW metric is inside a trapped surface $r_\Lambda \equiv 1/H_\Lambda= (8\pi G\rho_\Lambda/3)^{-1/2}$, which behaves like the interior of a BH. 
To see this, consider the maximum distance $r_*$ traveled by an
outgoing radial null geodesic (the Event Horizon at $\tau$, \cite{Ellis1993}) in the FLRW metric:
\beq
r_* \equiv a \chi_* = a \int_\tau^\infty \frac{d\tau}{a(\tau)} =
a \int_a^\infty \frac{d\ln{a}}{a H(a)} < \frac{1}{H_\Lambda} \equiv r_{\Lambda}
\label{eq:chi}
\eeq
where $\chi_*(a)$ is the corresponding comoving scale. 
For~small $a$,
 the value of
$\chi_*$ is fixed to a constant $\chi_* \simeq 3r_\Lambda$. Thus, the~physical trapped surface radius $r_*$ increases with time. 
As~we approach $a \simeq 1$,
 the Hubble rate becomes constant and $r_*$ freezes to a constant value $r_* = r_\Lambda$. 
 No signal from inside $r_*$ can reach the outside, just like in the interior of a BH. 
 In~fact, according to the Birkhoff theorem (see~\cite{Deser2005}), the~metric outside should be exactly that of the SW BH in the limit of an empty outside space. 
 So,
  the FLRW metric is a SW BH from the outside with $r_{S}=\rL$. 
  This breaks homogeneity (on scales larger than $\rL$), but~this is needed if we want causality.  Homogeneity is  inconsistent with a causal~origin. 

We next look for a solution to EFE with spherical symmetry in the SW frame where we have matter $\rho_m=\rho_m(t,r)$ and radiation $\rho_R=\rho_R(t,r)$ inside some radius $R$ and empty space outside:
\vspace{-6pt}
\bea
\rho(t,r) &=& \left\{ \begin{array}{ll} 
0  &  {\text{for}} ~ ~ r>R \\
 \rho_m + \rho_R& {\text{for}} ~ ~r<R \\
\end{array} \right. .
\label{eq:rhoVu}
\eea

The solution can be given in terms of the metric in Equation~(\ref{eq:newFRW}) with:
\vspace{-6pt}
\beq
2\Phi(t,r) = \left\{ \begin{array}{ll} 
  - r_{S}/r  &  {\text{for}} ~ ~ r>R\\
 -r^2 H^2(t,r)  & {\text{for}} ~ ~r<R \\
\end{array} \right.
\label{eq:BH.u2}
\eeq
where for $r>R$, we have the SW metric. 
At~the junction $r=R$, we find that:
\beq
R=[r_H^2 r_S]^{1/3}.
\label{eq:R}
\eeq

Using Equation~(\ref{eq:newFR3}), we can immediately see that the solution is  $H(t,r)=H(\tau)$
and $R(\tau)=[r_S/H^2(\tau)]^{1/3}$, where $\tau$ is the proper time of an observer comoving with the fluid.
Given $\rho(\tau)$ and $p(\tau)$ in the interior,
 we can find $H(\tau)$ and $R(\tau)$ as:
\beq
H^2(\tau) = \frac{8\pi G}{3} \rho(\tau) = \frac{r_S}{ R^3(\tau)}
\eeq

\begin{figure}
\includegraphics[width=.78\linewidth]{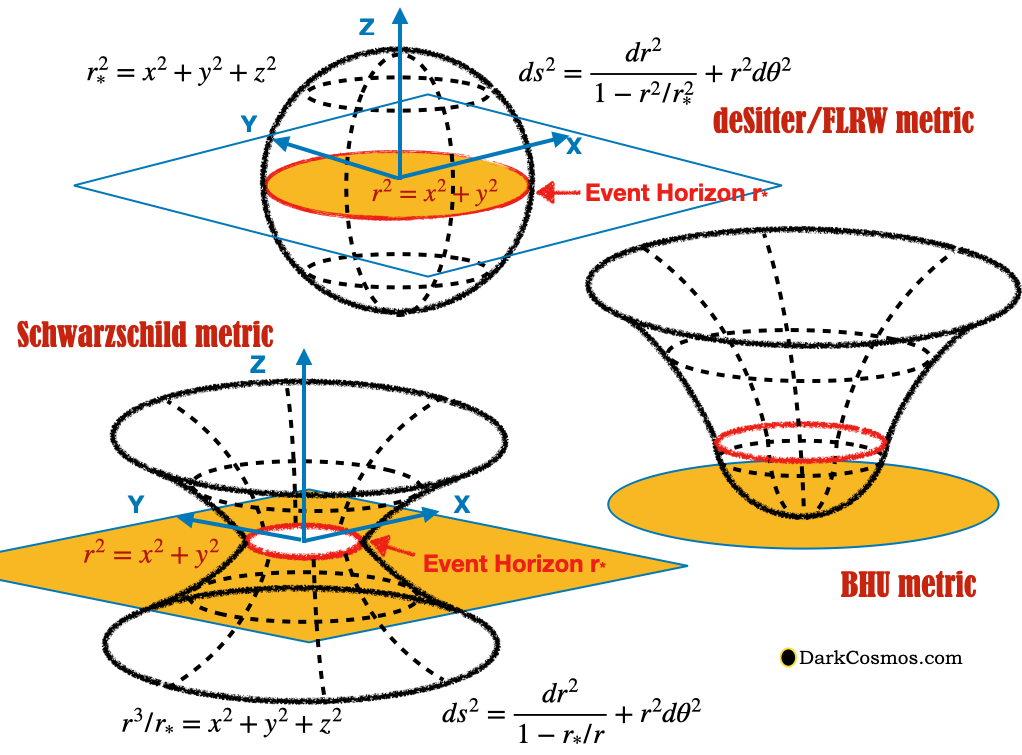}
\caption{Spatial 2D polar representation 
of $ds^2= (1+2\Phi)^{-1} dr^2 + r^2 d\theta^2$  metric embedded in a 3D flat space. deSitter space (top,  $2\Phi=-r^2/r_*^2$) corresponds to a 2D sphere (S2) in 3D flat space.
Yellow region shows the projected coverage in the true $(x,y)$-plane. 
Coordinate $z$ is an auxiliary extra dimension for illustration  ($z^2=r_*^2 -r^2$ for deSitter). 
The true 3D space has an additional angle in $d\Omega$, which is not shown.  
Schwarzschild metric (bottom left) corresponds to $2\Phi=-r_*/r$ and  the combined BHU metric is shown in the middle right.}
\label{fig:BHU}
\end{figure}

This corresponds to a homogeneous FLRW cloud of fixed mass $M=r_S/2G$ in free-fall inside $R(\tau)$. %PLEASE CHECK. 
For constant  $H(\tau)=H_{\Lambda} \equiv  r_\Lambda^{-1}$,
the FLRW metric becomes the dS metric; 
so, the interior solution becomes Equation~(\ref{eq:deSitter})
with $R=\rL=r_S=r_H$. 
Such solutions are illustrated in Figure~\ref{fig:BHU}. 

The  corresponding comoving radius to $R(\tau)$ is $ \chi_*(\tau) \equiv  R(\tau)/ a(\tau)$. We can see how $R$ can be related with a free-fall geodesic radial shell:
\beq
\frac{dR}{d\tau} = a \frac{d\chi_*}{d\tau} + \chi_* \frac{da}{d\tau}  
= V_0 + HR = V_0 + (r_S/R)^{1/2}
\label{eq:Rdot}
\eeq
where $V_0\equiv a\dot{\chi}_*$.  For~a time-like geodesic of constant $\chi_*$ ($d\chi=0$),
 we have  $V_0=0$. 
For~a null-like geodesic ($ad\chi=d\tau$), $V_0=1$. 
This  shows that a solution for $R(\tau)$ exists for any content inside $R$.
Israel junction conditions~\cite{Israel} for continuity of the metric and their derivative (the extrinsic curvature) are fulfilled for Equation~(\ref{eq:R}) with no surface terms 
(see~\cite{hal-03344159,universe}). %PLEASE CHECK.

\begin{figure}
\includegraphics[width=0.95\linewidth]{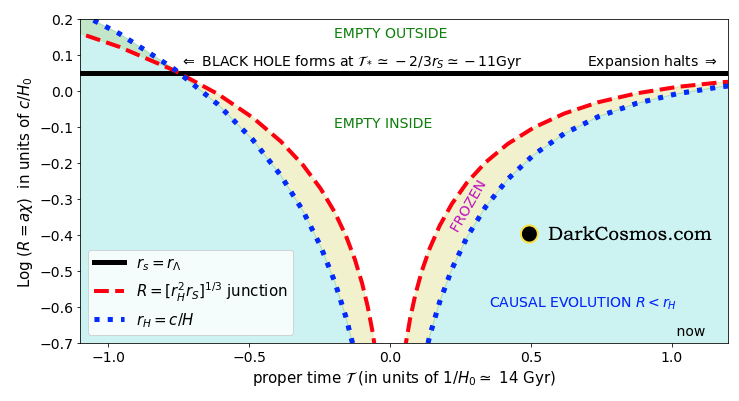}
\caption{Physical
 radius $R$ of a homogeneous spherical cloud as a function of proper (or comoving) time $\tau$. A~cloud of radius $R$ and mass $M$ collapses in free-fall. When it reaches $R=r_{S} =2GM$, at $\tau_* \simeq 2r_S/3 \simeq -11$Gyr,  it becomes a BH.  The~collapse proceeds inside the BH until it bounces producing an expansion (the hot Big Bang). The~event horizon $r_{S}$ behaves like a cosmological constant with $\Lambda= 3/r_{S}^2$ so that the expansion freezes before it reaches back to $r_{S}=\rL$. 
The calculation corresponds to $\Omega_\Lambda=0.7$ with  $ R=[r_H^2 r_{S}]^{1/3}$. 
Structure in between $R$ and $r_H$ is frozen (yellow shading) and seeds structure formation in our  Universe.
Blue shading indicates causal evolution ($R<r_H$).}
\label{fig:BHevolution}
\end{figure}

Note that for $R<r_S$ (i.e., inside the BH),  Equation~(\ref{eq:R})  indicates that there is a region with no matter, $r_S>r>R$, and a region with matter outside the Hubble horizon, $R>r>r_H$.  %PLEASE CHECK.
This region is dynamically frozen because  the time a perturbation takes to travel that distance is larger than the expansion time~\cite{Dodelson}.
The frozen region increases during collapse inside $r_S$ and re-enters $r_H$ during expansion. %PLEASE CHECK.
This is a potential source of frozen perturbations, which acts very much like Cosmic Inflation. 
This is illustrated by a yellow shading in Figure~\ref{fig:BHevolution}.

%%%%%%%%%%%%%%%%%%%%%%%%%%%%%%%%%%%%%%%%%%%%%
\subsection{The GHY Boundary~Term} 
\label{sec:GHY}

The expansion that happens inside an isolated BH is bounded by its event horizon $r<r_S$ and we need to add the GHY boundary term  $S_{GHY}$ to the action in Equation~(\ref{eq:actionK}),~where:
\beq
S_{GHY}= \frac{1}{8\pi G} \oint_{\partial \calMa}   d^3y \sqrt{-h} \,K
\label{eq:actionGHY}
\eeq

The integral is over the induced metric at ${\partial \calMa}$, which for a time-like junction $d\chi=0$
corresponds to $R=r_S$:
\beq
ds^2_{{\partial \calMa}}= h_{\alpha\beta}  dy^\alpha dy^\beta= -d\tau^2 + r_S^2 \dA^2
\label{eq:SigmaR}
\eeq

So, the only remaining degrees of freedom in the action are time $\tau$ and the angular coordinates.
We can use this metric and the trace of the extrinsic curvature at $R=r_S$
to estimate $K = -2/r_S$  \cite{hal-03344159}. 
This result is also valid for a null geodesic. 
We then have:
\beq
S_{GHY}= \frac{1}{8\pi G} \int d\tau \, 4\pi r_S^2 \, K = - \frac{r_S}{G} \tau
\label{eq:actionGHY2}
\eeq

The $\Lambda$ contribution to the action in   Equation~(\ref{eq:actionK}) is:
$S_{\Lambda}= - \Lambda \calMa/(8\pi G) =   - r_S^3 \Lambda \tau /3G $
where we have estimated the total 4D volume $ \calMa $ as that bounded by  $\partial \calMa$ inside $r<r_S$, i.e., $ \calMa = 2V_3\tau$, where the factor 2 accounts for the fact that  $V_3= 4\pi r_S^3/3 $  is covered twice, first during collapse and again during expansion. 
Comparing the two terms, we can see that we need $\Lambda =3 r_S^{-2} $ or, 
equivalently, $\rL=r_S$ to cancel the boundary term. 
In~other words,
the  evolution inside a BH event horizon induces a $\Lambda$ term in the EFE even when there is no $\Lambda$ term to start with.
Such an event horizon becomes a boundary for outgoing geodesics, i.e.,~expanding solutions. 
This provides a fundamental interpretation to the observed $\Lambda$ as a causal boundary~\cite{Gaztanaga2020,Gaztanaga2021}.

%%%%%%%%%%%%%%%%%%%%%%%%%%%%
\section{How \boldmath{$\Lambda$} Emerges}
\label{sec:evolution}

We will start assuming that $\Lambda=0$ and show how $\Lambda$ emerges because of the dynamics inside the BH event horizon.
Consider a large cloud dominated by dust ($\rho_m$ with $p=0$) with radius $R$ and  mass $M$, surrounded by a region of empty space.  This is a good approximation for our Universe with $M \simeq 5 \times 10^{22} M_{\odot}$,  which corresponds to the mass inside our future FLRW event horizon $\rL$ in Equation~(\ref{eq:chi}).
The resulting BH density is very low, $\rho_{BH} = 3 r_S^{-2}/8\pi G= \rho_\Lambda$, which is 25\% lower than the critical density today  $\rho = 3 H_0^{2}/8\pi G$ (a few protons per cubic meter). 

Gravity will make such a dust cloud 
collapse following a free-fall time-like geodesic of Equation~(\ref{eq:R}):
$R=(r_H^2 r_S)^{1/3}$. 
As there is no pressure support,  it will eventually collapse into a BH  when $R=r_{S} =r_H= 2GM$.
This happened 25~Gyrs ago according to the numerical result in Figure~\ref{fig:BHevolution}.
The collapse continues free-fall inside  the BH as illustrated in the left half of Figure~\ref{fig:BHevolution}.
 
As the density inside gets larger, matter ionizes and radiation
dominates the collapse, as it happens in stars. After~11Gyrs, the~density reaches the density of a neutron star. 
The collapse is halted by neutron degeneracy,
causing the implosion to rebound and bounce back.
This happens well before Planck times (Quantum Gravity).
The Hubble radius at that time corresponds to a solar mass, so the situation is similar to the interior of a collapsing star.  This should lead to a bounce, either because of supernova explosion or because of the elasticity of the neutron density \cite{elasticity}. This Big Bounce generates a hot Big Bang, similar to the bouncing of a ball on the floor \cite{universe}.

The~resulting radiation and plasma fluids
cool down as they expand into the standard FLRW evolution: Nucleosynthesis and Cosmic Microwave Background (CMB) decoupling.  Frozen perturbations from outside $r_H$ re-enter the horizon as the expansion proceeds. These are the seeds for new structure (Baryon Acoustic Oscillations or BAO and galaxies) that grow under gravitational instability. Finally, the~expansion freezes (exponential inflation) as it reaches back to the original BH event horizon $R=r_{S}$.  As~detailed in Section~\ref{sec:GHY}, the~expansion inside a BH event horizon induces a $\Lambda$ term in EFE, even when there is none to start with. 
 The~expansion comes to a halt in the physical frame as nothing can come out of a BH. This provides a fundamental interpretation to the observed $\Lambda$ as causal boundary~\cite{Gaztanaga2020,Gaztanaga2021} and also for the origin of the Big~Bang.

%%%%%%%%%%%%%%%%%%%%%%%%%%%%%%%%%%%%%%%%%% 
%%%%%%%%%%%%%%%%%%%%%%%%%%%%%%%%%%%%%%%%%%%%%
%%%%%%%%%%%%%%%%%%%%%%%%%%%%%%%%%%%%%%%%%%%%%
\section{Discussion and Conclusions}
\label{sec:conclusion}

%%%%%%%%%%%%%%%%%%%%%%%%%%%%
%\subsection{Physical BHs}

The observed cosmic acceleration directly implies that the FLRW metric has an Event Horizon  $r_S=\rL$ in Equation~(\ref{eq:chi}), which agrees well with the BHU idea~\cite{hal-03344159,universe}. Moreover,
the measured cosmic density $\rho$ and expansion rate also match well those of a BH 
with $r_S=\rL=2GM$ and BH density $\rho= 3r_S^{-2}/(8\pi G)$.
Thus,
 the BHU provides a fundamental explanation to the meaning and origin of the $\Lambda$ term
and the hot Big Bang, as~shown in Figure~\ref{fig:BHevolution}.

Inflation (\cite{Starobinski1979,Guth1981,Linde1982,Albrecht1982})
is an important ingredient in the standard cosmological model. 
For a review,
 see~\cite{Liddle1999,Dodelson}.
It solves several problems, the~most relevant here are:
(1) the horizon problem;
(2) the source of structure in the universe (such BAO);
(3) the flatness problem.
The horizon and structure problems arise because much of the structure that we observe today, e.g.,~BAO in CMB maps, was outside the Hubble horizon $r_H$ at the time of light emission and therefore could not have had a causal origin if there was nothing before the Big Bang. 
Inflation solves all  these problem at the prize of introducing several new ad~hoc  ingredients to the early universe that are hard to~test. %PLEASE CHECK.

As shown by the yellow shaded region in  Figure~\ref{fig:BHevolution}, a~large fraction of the mass $M$ that collapsed into our BHU is outside $r_H$. This solves the horizon problem. The~same mass outside $r_H$ can be the source of structure as it re-enters $r_H$.  Harrison~\cite{Harrison1970}  and Zel'dovich~\cite{Zeldovich1970} independently proposed in 1970's (see also~\cite{PeeblesYu}) that gravitational instability of regular matter alone can generate a scale-invariant spectrum  of fluctuations, very similar to that produced by models of~Inflation.

We have assumed here a global flat topology $k=\Lambda=0$ because this reproduces
the  Minkowski metric for empty space.
However,~the BHU solutions are also exact GR solutions for $k\ne \Lambda \ne 0$ by just replacing $r_H^{-2} \equiv H^2 + k/a^2 + \Lambda/3$ in  Equation~(\ref{eq:R}). 
Given some energy content, EFE cannot be used to decide the topology $k$ of the metric or the value of the effective $\Lambda$. 
This is a global property that is either assumed or directly measured. 
So, any choice other than $\Lambda=k=0$ would require some justification that is outside  GR. In~that respect, we believe there is no flatness or $\Lambda$ problem within~GR.
 
At the time of CMB last scattering, $R$ corresponds to an angle $\theta \simeq 60$ deg. 
So,
 we can actually observe perturbations larger than $R$. 
 Scales that are not causally connected! 
 There is good observational evidence for homogeneity and lack of correlations in the CMB at $r > R$ (see~\cite{Benjamin} and references therein).
This could be related to the so-called CMB anomalies (i.e., apparent deviations with respect to simple predictions from $\Lambda$ Cold Dark Matter, see~\cite{Gaztanaga2021,FG20,GF21} and references therein), or~the apparent tensions in measurements from vastly different cosmic scales or times (e.g.,~\cite{P18cosmo,Riess19,DES2021,DiValentino}). 
This requires further investigation that will be presented in a separate~publication \cite{universe}.

\acknowledgments{This work has been supported by Spanish MINECO  grants PGC2018-102021-B-100 and EU grants LACEGAL 734374 and EWC 776247 with ERDF funds. 
IEEC is funded by Generalitat de~Catalunya.}%In this section you can acknowledge any support given which is not covered by the author contribution or funding sections. This may include administrative and technical support, or donations in kind (e.g., materials used for experiments).

\conflictsofinterest{The author declares no conflict of~interest.}

%\begin{adjustwidth}{-\extralength}{0cm}
%\centering %% If there is a figure in wide page, please release command \centering
\reftitle{References}

\end{paracol}

\begin{thebibliography}{999}

\bibitem[{Weinberg}(1989)]{Weinberg1989}
{Weinberg}, S.
\newblock {The cosmological constant problem}.
\newblock {\em Rev. Mod. Phys.} {\bf 1989}, {\em 61},~1--23. https://doi.org/10.1103/RevModPhys.61.1.

\bibitem[{Carroll} \em{et~al.}(1992){Carroll}, {Press}, and
  {Turner}]{Carroll1992}
{Carroll}, S.M.; {Press}, W.H.; {Turner}, E.L.
\newblock {The cosmological constant.}
\newblock {\em Annu. Rev. Astron. Astrophys.} {\bf 1992}, {\em 30},~499--542. https://doi.org/10.1146/annurev.aa.30.090192.002435.

\bibitem[{Peebles} and {Ratra}(2003)]{Peebles2003}
{Peebles}, P.J.; {Ratra}, B.
\newblock {The cosmological constant and dark energy}.
\newblock {\em Rev. Mod. Phys.} {\bf 2003}, {\em 75},~559--606. https://doi.org/10.1103/\linebreak RevModPhys.75.559.

\bibitem[{Einstein}(1916)]{Einstein1916}
{Einstein}, A.
\newblock {Die Grundlage der allgemeinen Relativit{\"a}tstheorie}.
\newblock {\em Ann. Phys.} {\bf 1916}, {\em 354},~769--822. 

\bibitem[{Einstein}(1917)]{Einstein1917}
{Einstein}, A.
\newblock {Kosmologische Betrachtungen zur allgemeinen
  Relativit{\"a}ts-Theorie}.
\newblock In {\em Das Relativitätsprinzip}; Vieweg+ Teubner Verlag: Wiesbaden, Germany, {1917}; pp. {142--152.} 


\bibitem[Elizalde()]{Elizalde2021}
Elizalde, E.
\newblock \emph{The True Story of Modern Cosmology: Origins, Main Actors and  Breakthroughs}; Springer Nature: Berlin/Heidelberg, Germany, 2021.


\bibitem[{Gazta{\~n}aga}(2021)]{Gaztanaga2021}
{Gazta{\~n}aga}, E.
\newblock {The cosmological constant as a zero action boundary}.
\newblock {\em Mon. Not. R. Astron. Soc.} {\bf 2021}, {\em 502},~436--444. https://doi.org/10.1093/mnras/stab056.

\bibitem[{Padmanabhan}(2010)]{Padmanabhan}
{Padmanabhan}, T.
\newblock {\em {Gravitation}}; Cambridge University Press: Cambridge, UK, 2010.

\bibitem[{Calder} and {Lahav}(2008)]{CalderLahav2008}
{Calder}, L.; {Lahav}, O.
\newblock {Dark energy: Back to Newton?}
\newblock {\em Astron. Geophys.} {\bf 2008}, {\em 49},~1.13--1.18. https://doi.org/10.1111/j.1468-4004.2008.49113.x.

\bibitem[{Gazta\~naga}(2020)]{Gaztanaga2020}
{Gazta\~naga}, E.
\newblock {The size of our causal Universe}.
\newblock {\em Mon. Not. R. Astron. Soc.} {\bf 2020}, {\em 494},~2766--2772. https://doi.org/10.1093/\linebreak mnras/staa1000.

\bibitem[{O'Raifeartaigh} and {Mitton}(2015)]{SteadyState}
{O'Raifeartaigh}, C.; {Mitton}, S.
\newblock {A new perspective on steady-state cosmology}. \emph{arXiv} {\bf 2015}, 
\newblock  arXiv:1506.01651.

\bibitem[{Bondi} and {Gold}(1948)]{SSC1}
{Bondi}, H.; {Gold}, T.
\newblock {The Steady-State Theory of the Expanding Universe}.
\newblock {\em Mon. Not. R. Astron. Soc.} {\bf 1948}, {\em 108},~252. https://doi.org/10.1093/mnras/108.3.252.

\bibitem[{Hoyle}(1948)]{SSC2}
{Hoyle}, F.
\newblock {A New Model for the Expanding Universe}.
\newblock {\em Mon. Not. R. Astron. Soc.} {\bf 1948}, {\em 108},~372. https://doi.org/10.1093/mnras/\linebreak 108.5.372.

\bibitem[{Dodelson}(2003)]{Dodelson}
{Dodelson}, S.
\newblock {\em {Modern Cosmology}}; Academic Press: New York, NY, USA,  2003.

\bibitem[Gaztanaga(2021)]{hal-03344159}
Gaztanaga, E.
\newblock {The Black Hole Universe (BHU) from a FLRW Cloud}.
\newblock Submitted to Physics of the Dark Universe.
 Available as preprint:   
 \url{https://hal.archives-ouvertes.fr/hal-03344159} 
 accessed on September 14, 2021.
 
 \bibitem[Gaztanaga(2022)]{universe}
Gaztanaga, E.
\newblock {How the Big Bang End Up Inside a Black Hole}.
\newblock {\em Universe} Available as preprint:
\url{https://www.preprints.org/manuscript/202201.0459/v1} 
 accessed on Feb 1, 2022.
 

\bibitem[{Bera} \em{et~al.}(2020){Bera}, {Jones}, and {Andersson}]{elasticity}
{Bera}, P.; {Jones}, D.I.; {Andersson}, N.
\newblock {Does elasticity stabilize a magnetic neutron star?}
\newblock {\em MNRAS} {\bf 2020}, {\em 499},~2636--2647.

\bibitem[{Mitra}(2012)]{Mitra2012}
{Mitra}, A.
\newblock {Interpretational conflicts between the static and non-static forms
  of the de Sitter metric}.
\newblock {\em Nat. Sci. Rep.} {\bf 2012}, {\em 2},~923. https://doi.org/10.1038/srep00923.

\bibitem[{Planck Collaboration}(2020)]{P18cosmo}
{Planck Collaboration}.
\newblock {Planck 2018 results. VI. Cosmological parameters}.
\newblock {\em Astron. Astrophys.} {\bf 2020}, {\em 641},~A6. https://doi.org/\linebreak 10.1051/0004-6361/201833910.

\bibitem[{DES Collaboration}(2021)]{DES2021}
{DES Collaboration}.
\newblock Dark Energy Survey Year 3 Results: Cosmological Constraints from  Galaxy Clustering and Weak Lensing. {\em arXiv} {\bf 2021}, arXiv:2105.13549.

\bibitem[{Starobinski{\v{i}}}(1979)]{Starobinski1979}
{Starobinski{\v{i}}}, A.A.
\newblock {Spectrum of relict gravitational radiation and the early state of
  the universe}.
\newblock {\em JETP Lett.} {\bf 1979}, {\em
  30},~682--685.

\bibitem[{Guth}(1981)]{Guth1981}
{Guth}, A.H.
\newblock {Inflationary universe: A possible solution to the horizon and
  flatness problems}.
\newblock {\em Phys. Rev. D} {\bf 1981}, {\em 23},~347--356. https://doi.org/10.1103/PhysRevD.23.347.

\bibitem[{Linde}(1982)]{Linde1982}
{Linde}, A.D.
\newblock {A new inflationary universe scenario}.
\newblock {\em Phys. Lett. B} {\bf 1982}, {\em 108},~389--393. https://doi.org/10.1016/0370-2693(82)91219-9.

\bibitem[{Albrecht} and {Steinhardt}(1982)]{Albrecht1982}
{Albrecht}, A.; {Steinhardt}, P.J.
\newblock {Cosmology for Grand Unified Theories with Radiatively Induced
  Symmetry Breaking}.
\newblock {\em Phys. Rev. Lett.} {\bf 1982}, {\em 48},~1220--1223. https://doi.org/10.1103/PhysRevLett.48.1220.

\bibitem[{Hilbert}(1915)]{Hilbert1915}
{Hilbert}, D.
\newblock {Die Grundlage der Physick}.
\newblock {\em Konigl. Gesell. D Wiss. Gött. Math.-Phys. K} {\bf 1915},
  {\em 3},~395--407. 

\bibitem[{Landau} and {Lifshitz}(1971)]{Landau1971}
{Landau}, L.D.; {Lifshitz}, E.M.
\newblock {\em {The Classical Theory of Fields}};  {Pergamon Press:}, NY, 1971.


\bibitem[{Carroll}(2004)]{Carroll2004}
{Carroll}, S.M.
\newblock {\em {Spacetime and Geometry}}; {Cambridge U.Press:}, Cambridge
 2004.

\bibitem[{York}(1972)]{York}
{York}, J.W.
\newblock {Role of Conformal Three-Geometry in the Dynamics of Gravitation}.
\newblock {\em Phys. Rev. Lett.} {\bf 1972}, {\em 28},~1082--1085. https://doi.org/10.1103/PhysRevLett.28.1082.

\bibitem[{Gibbons} and {Hawking}(1977)]{Gibbons}
{Gibbons}, G.W.; {Hawking}, S.W.
\newblock {Cosmological event horizons, thermodynamics, and particle creation}.
\newblock {\em Phys. Rev. D} {\bf 1977}, {\em 15},~2738--2751. https://doi.org/10.1103/PhysRevD.15.2738.

\bibitem[{Hawking} and {Horowitz}(1996)]{Hawking1996}
{Hawking}, S.W.; {Horowitz}, G.T.
\newblock {The gravitational Hamiltonian, action, entropy and surface terms}.
\newblock {\em Class. Quantum Gravity} {\bf 1996}, {\em 13},~1487--1498. https://doi.org/10.1088/0264-9381/13/6/017.

\bibitem[{Garc{\'\i}a-Bellido} and {Espinosa-Portal{\'e}s}(2021)]{GB2021a}
{Garc{\'\i}a-Bellido}, J.; {Espinosa-Portal{\'e}s}, L.
\newblock {Cosmic acceleration from first principles}.
\newblock {\em Phys. Dark Univ.} {\bf 2021}, {\em 34},~100892. https://doi.org/10.1016/j.dark.2021.100892.

\bibitem[{Espinosa-Portal{\'e}s} and {Garc{\'\i}a-Bellido}(2021)]{GB2021b}
{Espinosa-Portal{\'e}s}, L.; {Garc{\'\i}a-Bellido}, J.
\newblock {Covariant formulation of non-equilibrium thermodynamics in General
  Relativity}.
\newblock {\em Phys. Dark Universe} {\bf 2021}, {\em 34},~100893. https://doi.org/10.1016/j.dark.2021.100893.

\bibitem[{Arjona} \em{et~al.}(2021){Arjona}, {Espinosa-Portales},
  {Garc{\'\i}a-Bellido}, and {Nesseris}]{GB2021c}
{Arjona}, R.; {Espinosa-Portales}, L.; {Garc{\'\i}a-Bellido}, J.; {Nesseris},
  S.
\newblock {A GREAT model comparison against the cosmological constant}.
\newblock {\em arXiv} {\bf 2021}, arXiv:2111.13083.

\bibitem[{Ellis} and {Rothman}(1993)]{Ellis1993}
{Ellis}, G.F.R.; {Rothman}, T.
\newblock {Lost horizons}.
\newblock {\em Am. J. Phys.} {\bf 1993}, {\em 61},~883--893. https://doi.org/10.1119/1.17400.

\bibitem[{Deser} and {Franklin}(2005)]{Deser2005}
{Deser}, S.; {Franklin}, J.
\newblock {Schwarzschild and Birkhoff a la Weyl}.
\newblock {\em Am. J. Phys.} {\bf 2005}, {\em 73},~261--264. https://doi.org/10.1119/1.1830505.

\bibitem[{Israel}(1967)]{Israel}
{Israel}, W.
\newblock {Singular hypersurfaces and thin shells in general relativity}.
\newblock {\em Nuovo Cimento B Ser.} {\bf 1967}, {\em 48},~463--463. https://doi.org/10.1007/BF02712210.

\bibitem[{Liddle}(1999)]{Liddle1999}
{Liddle}, A.R.
\newblock {Observational tests of inflation}.
\newblock {\em arXiv} {\bf 1999}, arXiv:astro-ph/astro-ph/9910110.

\bibitem[{Harrison}(1970)]{Harrison1970}
{Harrison}, E.R.
\newblock {Fluctuations at the Threshold of Classical Cosmology}.
\newblock {\em Phys. Rev. D} {\bf 1970}, {\em 1},~2726--2730. https://doi.org/10.1103/\linebreak PhysRevD.1.2726.

\bibitem[{Zel'Dovich}(1970)]{Zeldovich1970}
{Zel'Dovich}, Y.B.
\newblock {Reprint of 1970A\&A.....5...84Z. Gravitational instability: An
  approximate theory for large density perturbations.}
\newblock {\em Astron. Astrophys.} {\bf 1970}, {\em 500},~13--18.

\bibitem[{Peebles} and {Yu}(1970)]{PeeblesYu}
{Peebles}, P.J.E.; {Yu}, J.T.
\newblock {Primeval Adiabatic Perturbation in an Expanding Universe}.
\newblock {\em  Astrophys. J.} {\bf 1970}, {\em 162},~815. https://doi.org/10.1086/150713.

\bibitem[{Camacho} and {Gazta{\~n}aga}(2021)]{Benjamin}
{Camacho}, B.; {Gazta{\~n}aga}, E.
\newblock {A measurement of the scale of homogeneity in the Early Universe}.
\newblock {\em arXiv} {\bf 2021}, arXiv:2106.14303.

\bibitem[{Fosalba} and {Gazta{\~n}aga}(2021)]{FG20}
{Fosalba}, P.; {Gazta{\~n}aga}, E.
\newblock {Explaining cosmological anisotropy: Evidence for causal horizons
  from CMB data}.
\newblock {\em Mon. Not. R. Astron. Soc.} {\bf 2021}, {\em 504},~5840--5862. https://doi.org/10.1093/mnras/stab1193.

\bibitem[{Gazta\~naga} and {Fosalba}(2021)]{GF21}
{Gazta\~naga}, E.; {Fosalba}, P.
\newblock {A peek outside our Universe}. \emph{arXiv} {\bf 2021},
\newblock  arXiv:2104.00521.

\bibitem[{Riess}(2019)]{Riess19}
{Riess}, A.G.
\newblock {The expansion of the Universe is faster than expected}.
\newblock {\em Nat. Rev. Phys.} {\bf 2019}, {\em 2},~10--12. https://doi.org/10.1038/s42254-019-0137-0.

\bibitem[{Di Valentino} \em{et~al.}(2021){Di Valentino}, {Mena}, {Pan},
  {Visinelli}, {Yang}, {Melchiorri}, {Mota}, {Riess}, and {Silk}]{DiValentino}
{Di Valentino}, E.; {Mena}, O.; {Pan}, S.; {Visinelli}, L.; {Yang}, W.;
  {Melchiorri}, A.; {Mota}, D.F.; {Riess}, A.G.; {Silk}, J.
\newblock {In the realm of the Hubble tension-a review of solutions}.
\newblock {\em Class. Quantum Gravity} {\bf 2021}, {\em 38},~153001. https://doi.org/10.1088/1361-6382/ac086d.

\end{thebibliography}
\end{document}